\documentclass[5p]{elsarticle}

\usepackage{lineno}
\usepackage[]{hyperref}
\usepackage{amsmath,amssymb}
\usepackage{xcolor,cuted,caption}
\usepackage{slashed,cancel}
\usepackage{lipsum}
\usepackage{braket}
\usepackage{multirow}
\usepackage{tikz}
\usetikzlibrary{patterns}
\usetikzlibrary{decorations.pathmorphing}
\modulolinenumbers[5]

\journal{Journal of \LaTeX\ Templates}

\makeatletter
\def\ps@pprintTitle{%
\let\@oddhead\@empty
\let\@evenhead\@empty
\def\@oddfoot{}
\def\@evenfoot{}}
\makeatother

\begin{document}

\begin{frontmatter}

\title{Helicity Constraints To Soft Factor Of All Spin}

\author{Andriniaina Narindra Rasoanaivo}
\ead{andriniaina@aims.ac.za}
\address{Sciences Exp\'erimentales et des Math\'ematiques, Ecole Normale Sup\'erieure,\\  Complexe Scolaire Ampefiloha BP 881, Universit\'e d'Antananarivo - Madagascar}




\begin{abstract}
In this note we derive a set of non-perturbative constraints for soft operators from the Wigner's little group scaling of scattering amplitudes in a soft limit. We also show that the resolution of such constraints generates a master formula for the analytic expression of the single soft factor of any given spin and helicity.
\end{abstract}


\end{frontmatter}


\section{Introduction}

In quantum field theory, soft theorems are used to factorize the low energy contributions from the high energy part of scattering amplitudes. In \cite{Weinberg:1965nx}, such theorem is used to show that the effect of attaching a soft particle to a scattering amplitude is the same as supplying a factor to the amplitude, soft factors.

Over the past couple of decades, it has become increasingly
clear that soft factorization is a universal property of scattering amplitudes, and the corresponding soft factors are operators and depend mainly on the asymptotic properties of the external particles involved in the process but not on the underline high energetic interactions. In \cite{Strominger:2017zoo}, it is suggested that such universality of the theorem could be understood in terms of asymptotic symmetries of the amplitudes and the physical properties of the soft particle. In \cite{Marotta:2019cip, Elvang:2016qvq}, it is shown that the form of the soft operators depends on the spin of the soft particle.

Despite the differences between the spin properties of photons and gravitons, their respective soft factors exhibit multiple similarities. In \cite{Weinberg:1965nx}, Weinberg showed that the leading soft factor for gravitons behaves simiarly to that for photons. Both contain the respective information on the corresponding soft particle, and are singular in the soft momentum region with a pole of order one. In \cite{Cachazo:2014fwa}, Cachazo and Strominger demonstrated that sub-leading corrections for soft graviton share the same structure as the sub-leading soft photon in \cite{Low:1958sn}. These sub-leading corrections are both constructed from the total angular momentum operator of the asymptotic particle states.
 
Regarding the deeper understanding of the factorization, much works have been done, revealing similar behaviors in a various theories: soft massless scalars \cite{ Marotta:2019cip, Campiglia:2017dpg}, soft photons \cite{Low:1958sn, Ozeren:2005mp,Lysov:2014csa}, and soft gravitons \cite{Weinberg:1965nx,Cachazo:2014fwa, Pasterski:2015tva}; in a various dimensions \cite{Afkhami-Jeddi:2014fia,Schwab:2014xua, Kalousios:2014uva}; and for multiple soft emissions \cite{AtulBhatkar:2018kfi,Chakrabarti:2017ltl, Klose:2015xoa}.

In this work, we focus on deriving universal constraints for soft operators, analogous to how scattering amplitudes are constrained by the Wigner's little group \cite{Conde:2016vxs, Arkani-Hamed:2017jhn} via helicity operators. First, we derive these constraints by considering the action of soft operators to preserve helicity constraints on amplitudes. Second, we solve the helicity constraints to fix the kinematics of the single leading soft factor of any given particle and present the general expression for soft factors of arbitrary spin and helicity.

\section{Helicity constraints}
Scattering amplitudes, with massless particles only, are known to be a simultaneous eigenstate of the helicity operators $\{\hat{H}_{i}\}$  associated to the  asymptotic particles of the process. In the modern approach of amplitude calculation, the actions of the different helicity operators are combined to form multiple constraints to the scattering amplitude. Such helicity constraints are used to fix the kinematics of three-point amplitude up to a coupling constant, as shown in \cite{Conde:2016vxs, Arkani-Hamed:2017jhn}. Indeed the helicity constraints are a direct manifestation of Lorentz symmetry via Wigner's little group action on the asymptotic states \cite{Cachazo:2005ga}. In more compact form, massless amplitudes are constrained to satisfy the equations
\begin{equation}
\hat{H}_{i}\ket{\mathcal{M}_n}=h_i\ket{\mathcal{M}_n},
\label{helicity_amplitude_0}
\end{equation}
where $i=\{1,\ldots,n\}$ and $\ket{\mathcal{M}_n}$ represent the $n$-particle amplitude state with respective helicities $h_i$.

The objective of this section is to investigate the relations between helicity operators and soft operators. 
Such relations are expected to be able to fix the kinematics of the single soft contribution.
\newpage

Following Weinberg's approach \cite{Weinberg:1965nx}, attaching a soft particle, labeled by $r$, to an $n$-particles state is given by

\begin{equation}
\hat{S}_r\ket{\mathcal{M}_n}=\ket{\mathcal{M}_{n+1}^{(r)}},
\label{soft_action}
\end{equation}
where $\ket{\mathcal{M}_{n+1}^{(r)}}$ is the $(n+1)$-particle state. The action of the soft operator has to preserve the constraints \eqref{helicity_amplitude_0} so the $(n+1)$-state remains a simultaneous eigenstate of the set of helicity operators $\{H_I\}=\{\hat{H}_i\}\cup\{\hat{H}_r\}$  
\begin{equation}
\hat{H}_I\ket{\mathcal{M}_{n+1}^{(r)}}=h_I\ket{\mathcal{M}_{n+1}^{(r)}},\label{helicity_amplitude_1}
\end{equation}
for $I=\{1,\ldots,n,r\}$. The consistency of the three relations \eqref{helicity_amplitude_0}, \eqref{soft_action}, and \eqref{helicity_amplitude_1} leads to the relation
\begin{equation}
\begin{aligned}
 \left [\hat{H}_I,\hat{S}_r\right ]\ket{\mathcal{M}_n}+\hat{S}_r\hat{H}_I\ket{\mathcal{M}_n}=h_I\hat{S}_r\ket{\mathcal{M}_n}.
\end{aligned}
\label{consistency}
\end{equation}
Since the amplitude state $\ket{\mathcal{M}_n}$ does not have any information on the soft particle, where $\hat{H}_r\ket{\mathcal{M}_n}=0$. Additionally, $\left [\hat{H}_i,\hat{S}_r\right ]=0$ for hard particles. Thus, the soft operator must satisfy the commutation relations  
\begin{equation}
\left [\hat{H}_{I},\hat{S}_{r}\right ]=h_r\delta_{Ir}\hat{S}_{r}.
\label{constraint}
\end{equation}
Here the Kronecker delta, $\delta_{Ir}$, reflects the fact that soft operators carry only information about the soft particle. 
The commutation relations \eqref{constraint} are equivalent to the helicity constraints on the amplitudes which is the manifestation of the Lorentz symmetry to constrain soft operators. In the next section, we will show how such helicity constraints could be used to fix the Weinberg's soft factor up to coupling constant, the same way as three point amplitudes in \cite{Conde:2016vxs, Arkani-Hamed:2017jhn}.

\section{Weinberg's soft factors}
The main goal of this section is to solve the constraints \eqref{constraint} for the Weinberg's soft factor. As presented in \citep{Cachazo:2014fwa}, such soft factor is the leading term of the soft operator in the soft momentum expansion 
 \begin{equation}
 \hat{S}_r=S^{(0)}+S^{(1)}+ S^{(2)}\cdots .
 \label{soft_factorization}
 \end{equation}
To simplify the resolution of helicity constraints, we will use spinor helicity formalism. Through out this work, the soft particle will be labeled by $r$, while  the hard particles will be labeled by Arabic letters $\{i,j,\ldots\}$. Thus for the soft factor, the commutation relations \eqref{constraint} can be decomposed as 
\begin{equation} 
\left [\hat{H}_r,S^{(0)}\right ]=h_rS^{(0)}\quad \text{and}\quad \left [\hat{H}_i,S^{(0)}\right ]=0.
\label{commutation_0}
\end{equation}
\subsection{Spinor helicity formalism} 
In this formalism, standard four-momentum $p_\mu$ of massless particles will be represented by a pair of left and right handed spinors $\lambda_a$ and $\tilde{\lambda}_{\dot{a}}$, such that 
\begin{equation}
p_{\mu}p^\mu=0\Longleftrightarrow p_\mu\sigma^\mu_{a\dot{a}}=\lambda_{a}\tilde{\lambda}_{\dot{a}}.
\label{onshell_condition}
\end{equation}
The introduction of these spinor variables makes the on-shell condition $p^2=0$ trivial, since a pair of spinors $\lambda$ and $\tilde{\lambda}$ is associate to a null momentum up to Wigner's transformation \cite{Henn:2014yza}. 
Lorentz invariant quantity could be built from $\lambda$'s and $\tilde{\lambda}$'s through the standard angle and square brackets as basic spinor invariants
\begin{equation}
\braket{IJ}=\epsilon^{ab}\lambda_{a}^I\lambda_{b}^J\quad\text{and}\quad [IJ]=\epsilon^{\dot{a}\dot{b}}\tilde{\lambda}_{\dot{a}}^I\tilde{\lambda}_{\dot{b}}^J,
\end{equation}
where $\epsilon^{ab}$ and $\epsilon^{\dot{a}\dot{b}}$ are the antisymmetric two tensors. In terms of these angle and square brackets, the non-invariant scalar products may be written as
\begin{equation}
p^I\cdot p^J=\frac{1}{2}\braket{IJ}[IJ].
\label{scalar}
\end{equation}
Note that the Wigner's little group is a subgroup of the Lorentz that leaves a momentum $p^I_\mu$ invariant. In the spinor representation, Wigner's little group becomes a simple scaling transformations on the pair of spinors $\lambda^I$ and $\tilde{\lambda}^I$, and the corresponding generator, helicity operator, is nothing than a homogeneous dilatation operator acting on the spinor variables. As shown in \cite{Conde:2016vxs, Arkani-Hamed:2017jhn,Cachazo:2005ga, Henn:2014yza}, such helicity operator is explicitly given by 
\begin{equation}
\hat{H_I}=-\frac{1}{2}\left (\lambda_{a}^I\frac{\partial}{\partial \lambda_{a}^I}-\tilde{\lambda}_{\dot{a}}^I\frac{\partial}{\partial\tilde{\lambda}_{\dot{a}}^I}\right ).
\label{helicity_expression}
\end{equation}
The use of spinor variables tends to put the commutation relations \eqref{commutation_0} in a more convenient expression. Especially for the case of Weinberg's soft factor, the commutation relations are reduced into a set of  partial differential equations in $S^{(0)}$. The determination of the soft factor then reduces to solving corresponding PDE, parametrized by the helicity $h_r$. 

\subsection{Resolutions}

Before proceeding to the determination of the general expression of soft factors $S^{(0)}$, it is essential to set the different physical conditions that solution must satisfy. These conditions are used to separate physical solutions from the more general solutions of the commutation relations \eqref{commutation_0}.

First, from the conventional diagrammatic approach, the leading soft factor exhibits a universal infrared divergence as the momentum of the soft particle approaches zero \cite{Elvang:2016qvq,Cachazo:2014fwa,Low:1958sn}. This singularity originates from the propagator of an external hard particle with momentum \(p^i\) after emitting a soft massless particle with momentum \(q^r\). In the soft limit, the propagator reduces to \(1/(2p^i\cdot q^r)\), giving rise to the characteristic infrared pole \cite{Weinberg:1965nx}. To capture this universal behavior in our solution, we impose an \textit{asymptotic boundary condition} motivated by the same physical origin. At large momentum, the propagator contribution is suppressed, motivating the requirement that the soft factors $S^{(0)}$ vanish in the high-energy limit.

The second constraint follows from \textit{dimensional analysis} of scattering amplitudes. Since the leading soft theorem relates the \((n+1)\)-point and \(n\)-point amplitudes through relation \eqref{soft_action}, the mass dimension of the soft factor is fixed by the difference between the dimensions of the two amplitudes. After factoring out the coupling and charge factors, the mass dimension of the corresponding kinematic part of the soft factor is therefore fixed by the amplitude dimensions. For a spin-\(s\) soft particle, this dimensional constraint yields a mass dimension \(``s-2"\) for the kinematic of the soft factor, consistent with the familiar photon 
 and graviton 
  cases \cite{ Rodina:2018pcb, Broedel:2014fsa,Broedel:2014bza}. This requirement provides an additional constraint on the general solution and, together with the asymptotic boundary condition, allows the physically relevant soft factor to be identified.  

From \eqref{helicity_expression} it is easy to see that the set of partial differential equations associated to the helicity constraints are
\begin{equation}-\frac{1}{2}\left (\lambda_{a}^r\frac{\partial S^{(0)}}{\partial \lambda_{a}^r}-\tilde{\lambda}_{\dot{a}}^r\frac{\partial S^{(0)}}{\partial\tilde{\lambda}_{\dot{a}}^r}\right )=h_r S^{(0)},
\label{soft_pde_0}
\end{equation}
to describe the soft particle dependence of $S^{(0)}$, and 
\begin{equation}
-\frac{1}{2}\left (
\lambda_{a}^i\frac{\partial S^{(0)}}{\partial \lambda_{a}^i}-\tilde{\lambda}_{\dot{a}}^i\frac{\partial S^{(0)}}{\partial\tilde{\lambda}_{\dot{a}}^i}
\right )=0,
\label{hard_pde_0}
\end{equation}
for the hard kinematics dependence. For the resolution, it is more natural to solve the main equation \eqref{soft_pde_0} first and then use the equation \eqref{hard_pde_0} to adjust the kinematics of the solution of \eqref{soft_pde_0}. It is also necessary to use the method of separation variables in which $S^{(0)}$ is decomposed into products of two independent functions respectively depend on left and right handed spinor variables. Let $A$ be an holomorphic function that depends only on the variables $\lambda$ and respectively $\tilde{A}$ an anti-holomorphic function that depends only on the variables $\tilde{\lambda}$, such that 
\begin{equation}
S^{(0)}= A(\lambda^r)\;\tilde{A}(\tilde{\lambda}^r).
\label{separation}
\end{equation}
Taking the above expression into the equation \eqref{soft_pde_0} leads to the separation of variables  described by the two following partial differential equations respectively in $A$ and $\tilde{A}$
\begin{equation}\left \{\begin{aligned}
&\lambda_{a}^r\frac{\partial A}{\partial \lambda_{a}^r}=-(s+h_r)A\\
&\tilde{\lambda}_{\dot{a}}^r\frac{\partial \tilde{A}}{\partial\tilde{\lambda}_{\dot{a}}^r}=-(s-h_r)\tilde{A},
\end{aligned} \right .
\label{AA}
\end{equation}
here the separation parameter $s$ links the holomorphic and anti-holomorphic equations. In a non-invariant form, where all the spinor indexes are contracted, the generic solutions of the equation \eqref{AA} is given by
\begin{equation}
S^{(0)}=K(p^*\cdot q^r)
\underbrace{\braket{*r}^{-(s+h_r)} }_{A}\underbrace{[*r]^{-(s-h_r)}}_{\tilde{A}},
\label{generic_0}
\end{equation}
where $K$ is an arbitrary regular function invariant under the Wigner's little group transformation, while the star $``*"$ represents the kinematic variables  constructed from the hard momenta. To satisfy the equation \eqref{hard_pde_0} the above solution has to depend at least to two of momenta, which allow $K$ and the spinors ``$*$" to be fixed, leading to the solution
\begin{equation}
S^{(0)}= K(p^i\cdot q^r)\left (\frac{\braket{ij}}{\braket{ir}\braket{jr}}\right )^{\frac{s+h_r}{2}}
\left (\frac{[ij]}{[ir][jr]}\right )^{\frac{s-h_r}{2}}.
\label{generic_1}
\end{equation}
From the asymptotic boundary condition, $S^{(0)}$ has to vanish at high energy limit. Such condition can be expressed by considering the limit where $q^r$ tend to infinity, so the solution \eqref{generic_1} satisfies the asymptotic condition provided that $s+h_r$ and $s-h_r$ are positive, thus the following relation
\begin{equation}
-s\leq h_r\leq s,
\label{h_s}
\end{equation}
which implies the positivity of the separation parameter $s$. 
Even if the solution \eqref{generic_1} satisfies the asymptotic boundary condition by considering the relation \eqref{h_s}, still it needs to have the right mass dimension.
This dimension can be checked by expanding the function $K$ as
\begin{equation}
K=\sum_{\alpha\in\mathbb{Z}}
g_{\alpha+1}\left (\langle ir\rangle[ir]\right )^\alpha,
\label{expansion_serie}
\end{equation}
where \(\langle ir\rangle[ir]=2p^i\cdot q^r\) represents the scalar product between the soft momentum and the momentum of the \(i\)-th hard particle, as defined in \eqref{scalar}, and \(g_{\alpha+1}\) are coefficients. 

Since \(\langle ir\rangle [ir]\) has mass dimension two,
each kinematic term in the expansion \eqref{expansion_serie} has mass dimension $`` 2\alpha"$. The remaining the kinematic structure in equation \eqref{generic_1} contributes a factor of mass dimension \(``-s"\). Therefore, requiring the kinematic soft factor to have mass dimension \(``s-2"\) gives
\begin{equation}
2\alpha-s=s-2,
\end{equation}
and hence \(
\alpha=s-1.
\)

Here \(\alpha\) is an integer, then the dimensional constraint together with the relation \eqref{h_s} requires \(s\) to be a positive integer. This condition therefore allows the parameter \(s\) to be identified with the spin of the emitted soft particle.


The soft factor from the helicity constraints \eqref{commutation_0} are then parametrized by the spin and helicity of the soft particle. And since the soft particle can be emitted from any external hard leg ``$i$", the leading soft factor is obtained by summing the contributions from all hard external legs, each weighted by the appropriate factor $t^r_{i}$ describing its interaction with the soft particle.
\begin{flushleft}
\begin{equation}
\begin{aligned}
S^{(0)}_{s,h_r}=\sum_{i} \frac{g_s\; t^r_{i}}{\left (\braket{ir}[ir]\right )^{1-s}}
\left (\frac{\braket{ij}}{\braket{ir}\braket{jr}}\right )^{\frac{s+h_r}{2}}
\left (\frac{[ij]}{[ir][jr]}\right )^{\frac{s-h_r}{2}}.\\
\end{aligned}
\label{formula_0}
\end{equation}
\end{flushleft}
Here, the label $j$ is taken to be a reference spinor, as in \cite{Elvang:2016qvq,Cachazo:2014fwa,Arkani-Hamed:2008owk}, since it is used differently from $i$ in the expression above.
In the expression \eqref{formula_0}, $g_s$ represent the fundamental coupling associated to the soft spin-$s$ particle, and $t^r_{i}$ some factor associated to the charge of internal symmetry and signs for incoming/outgoing particles.

For $s=0$, the expression \eqref{formula_0} leads to the standard scalar soft factor for a cubic (Yukawa-like) interaction \cite{Briceno:2025ivl}, where  $t^r_i=1$ when the soft is emitted from an outgoing particle and $t^r_i=-1$ when it is emitted from an incoming,
\begin{flushleft}
\begin{equation}
\begin{aligned}
\\[-.4cm]
\hspace{-.2cm}S^{(0)}_{0,0}=g_0\sum_{i} \frac{\; t^r_{i}}{\braket{ir}[ir] }=g_0\sum_{i} \frac{t^r_{i}}{2p_i\cdot p_r }.
\end{aligned}
\label{s_0}
\end{equation}
\end{flushleft}

For $s=1$, the expression \eqref{formula_0} gives the following expressions
\begin{flushleft}
\begin{equation}\left \{
\begin{aligned}
&S^{(0)}_{1,-}=g_1\sum_{i} \; t^r_{i}
\frac{[ij]}{[ir][jr]},\\
&S^{(0)}_{1,+}=g_1\sum_{i} \; t^r_{i}
\frac{\braket{ij}}{\braket{ir}\braket{jr}}.
\end{aligned}
\right .
\label{s_1}
\end{equation}
\end{flushleft}
Here for the case of photon the $t_i^r=\pm 1$, depending on the emitting particle whether it is outgoing or incoming. While for the case of gluon, as shown in \cite{Dixon:2019lnw}, $t_i^r$ represent the color charge associated with the $i$-th particles from which the gluon is emitted and $j$ an arbitrary reference that represents a gauge choice in spinor helicity formalism. 

For $s=2$, the expression \eqref{formula_0} leads to standard soft factor for graviton as in \cite{Arkani-Hamed:2008owk} in which $t_i^r=\pm 1$, depending on the emitting particle whether it is outgoing or incoming, and $j$ an arbitrary reference spinor
\begin{flushleft}
\begin{equation}\left \{
\begin{aligned}
&S^{(0)}_{2,-}=g_2\sum_{i}\; t^r_{i}\frac{\braket{ir}[ij]^2}{[ir][jr]^2},\\
&S^{(0)}_{2,+}=g_2\sum_{i}\; t^r_{i}\frac{[ir]\braket{ij}^2}{\braket{ir}\braket{jr}^2}.
\end{aligned}\right .
\label{s_2}
\end{equation}
\end{flushleft}

\section{Discussion}

The main goal of this work is to develop universal constraints on soft operators. In the second section, we showed that these constraints are a direct manifestation of Wigner's little group and can be expressed as commutation relations between the soft operator and the helicity operators, as given in \eqref{commutation_0}. Using the spinor-helicity representation, the helicity operators reduce to differential operators, and each constraint ensures that the soft operator transforms according to the appropriate representation of Wigner's little group associated with the soft particle.

In the second part, we showed that, for the Weinberg soft factor, these constraints can be formulated as a set of partial differential equations. Solving these equations leads to an analytic expression for the soft factor parametrized by the helicity \(h_r\) of the soft particle and the separation parameter \(s\). To capture the infrared behavior of the soft factor, the parameter \(s\) is further constrained by the relation \eqref{h_s}, which allows \(s\) to be interpreted as the spin of the soft particle. Dimensional analysis then requires \(s\) to be an integer, implying that the soft factor obtained from our construction is associated with an integer-spin soft particle.

The result naturally includes the familiar cases of spin \(s=0\), \(s=1\), and \(s=2\), corresponding to scalar particles, gauge bosons, and gravitons, respectively, and formally extends the soft-emission structure to higher integer spins. However, the cases \(s\geq3\) require special consideration. In the soft limit, the coupling of a massless spin-\(s\) particle to hard external particles is constrained by the gauge invariance of the massless higher-spin field. In particular, Weinberg's consistency argument \cite{Weinberg:1964ew} shows that, for \(s\geq3\), the resulting constraint cannot in general be satisfied for arbitrary external momenta with nonzero couplings, which, under the assumptions of Weinberg's argument, requires the corresponding long-range higher-spin coupling to vanish, \(g_s=0\). Therefore, our generic solution should not be interpreted as establishing the existence of a nontrivial interacting massless higher-spin soft theorem. Rather, it characterizes the kinematic form that the soft factor would possess if such a coupling \(g_s\) were present, while the consistency of the coupling itself requires additional constraints beyond the kinematic analysis considered here.

These results show that the Wigner little-group constraints, together with dimensional analysis, determine the kinematic structure of the soft factors up to an overall coupling and charge factor, and reproduce the familiar soft factors for scalar particles, photons, and gravitons, see \cite{Weinberg:1965nx,Marotta:2019cip,Ozeren:2005mp}. In particular, the resulting soft factors take the expected universal form, with the remaining overall factor determined by the coupling of the soft particle to the corresponding hard external particles.


\bibliography{mybibfile}

\end{document}